\documentclass{elsarticle}

\usepackage{amsmath,amstext,amssymb}
\usepackage{subfigure,color}
\usepackage{psfrag,subfigure,color}
\usepackage{psfrag,subfigure,xcolor,color}
\usepackage{graphicx}
\usepackage{calc,bbm}

\def\ie{{i.\,e.\ }}
\def\eg{{e.\,g.\ }}

\def\re{\text{Re}\:}

\def\k{\kappa_5}

\def\U{\mathcal{U}}
\def\M{\mathcal{M}}

\def\del{\partial}
\def\diag{\mathrm{diag}}

\def\ee{{\mathrm e}}
\def\ii{{\mathrm i}}

\newcommand{\dd}{\mathrm{d}}

\begin{document}


\title{Black Hole Instability Induced by a Magnetic Field}
\date{\today}

\cortext[cor1]{Corresponding Author}

\author[UCLA]{Martin Ammon}
\ead{ammon@physics.ucla.edu}

\author[MPI]{Johanna Erdmenger}
\ead{jke@mppmu.mpg.de}
\author[MPI]{Patrick Kerner}
\ead{pkerner@mppmu.mpg.de}
\author[MPI]{Migael Strydom}
\ead{mstrydom@mppmu.mpg.de}

\address[UCLA]{Department of Physics and Astronomy\\University of California, Los Angeles, CA 90095, USA}
\address[MPI]{Max-Planck-Institut f\"ur Physik (Werner-Heisenberg-Institut)\\
 F\"ohringer Ring 6, 80805 M\"unchen, Germany}


\begin{abstract}
In the context of gauge/gravity duality, we find a new black hole
instability in asymptotically AdS spaces. On the field theory side, this instability is induced by a magnetic field in the vacuum, in contrast to previous instabilities which occur at finite density. On the gravity side, this corresponds to a spatial component of the gauge field in  $SU(2)$
Einstein-Yang-Mills theory, which provides the crucial non-Abelian
structure.   Our analysis may provide supporting evidence for recent QCD studies of $\rho$ meson condensation induced by a magnetic field.
\end{abstract}


\maketitle

\section{Introduction}
\label{sec:Introduction}

Within gauge/gravity duality,
asymptotically AdS black hole solutions and their stability properties
play a crucial role since black holes are dual to thermal states on
the field theory side. In particular, the stability of black holes is
connected to the stability of the thermal state. In recent years, it
was found that many asymptotically AdS black holes are unstable,
forming hair \cite{Gubser:2008zu,Gubser:2008px}. The most prominent
examples are charged AdS black holes which are unstable against
condensation of scalar \cite{Hartnoll:2008vx,Hartnoll:2008kx} or gauge fields \cite{Gubser:2008wv}. The condensate arising from a scalar or gauge field is a
holographic description of a superfluid with scalar or vector
condensates, respectively. 

In this letter we report on a new type of instability of Anti-de Sitter
black holes in $SU(2)$ Einstein-Yang-Mills theory. Our analysis is motivated
by recent field theory studies within QCD
\cite{Chernodub:2010qx,Chernodub:2010zw,Chernodub:2011mc,Chernodub:2011tv},
where  it has been proposed that a strong magnetic
field may generate $\rho$ meson condensation and 
superconductivity in the QCD vacuum\footnote{
This effect is similar to the Nielsen-Olsen instability for gluon condensation
\cite{Nielsen:1978rm} and to
$W$-boson condensation induced by a strong magnetic field 
\cite{Ambjorn:1988tm,Ambjorn:1989bd,Ambjorn:1989sz}}.
Essentially, in these papers
it has been argued that in a strong magnetic field, a gluon-mediated attraction
between quarks and antiquarks of different flavour leads to a colourless
spin-triplet bound state with quantum numbers of an electrically
charged $\rho^\pm$ meson. These quark-antiquark pairs condense to form
a new ground state.  Within gauge/gravity duality,
a similar result has been obtained in the Sakai-Sugimoto
model in \cite{Callebaut:2011uc,Callebaut:2011ab}, also at zero temperature as
in the field theory computation.

In this letter we find supporting evidence for this QCD proposal by
considering a suitable model within gauge/gravity duality at finite
temperature.  The
holographic model considered is $SU(2)$  Einstein-Yang-Mills theory in
$(4+1)$-dimensions. In the presence of a background $B$ field,
we do indeed find an instability, corresponding to the
formation of vector
hair in the gravity theory. In addition to the motivation from QCD
described above, the result is also of more general relevance in the context
of gravity and gauge/gravity duality. Whereas the QCD result has been obtained
at zero temperature, here we consider the finite temperature case.

In fact, the model we consider is a simple modification of the p-wave
holographic superconductor, first discussed in \cite{Gubser:2008wv},
in which there is a vector condensate in
the background of a chemical potential and density turned on by considering a
non-trivial radial profile for the temporal component $A_t$ of the
gauge field in the gravity theory. In addition, a p-wave superconductor
for which the dual field theory is explicitly known has been
constructed in \cite{Ammon:2008fc,Basu:2008bh,Ammon:2009fe} 
by embedding a probe
of two coincident D7-branes, corresponding to two flavours,
into the AdS-Schwarzschild black
hole background. This string-theoretical top-down approach lets one
identify the SU(2) chemical potential as an isospin chemical potential
and the condensate as a $\rho$ meson. 
A p-wave holographic model which includes the back-reaction of the
gauge field on the metric in the $SU(2)$ Einstein-Yang-Mills model has
been considered in \cite{Basu:2009vv,Ammon:2009xh}.

Here we show that a similar condensation occurs also when a
{\it spatial} component $A_y$
of the background $SU(2)$ gauge field, corresponding
to an external $B$ field,  has a non-trivial profile instead of the {\it
 temporal} component $A_t$ in the isospin chemical potential case.
We find
the corresponding holographic phase transition at finite
temperature. For this we
analyse the fluctuation quasinormal modes and find that they display an
instability. We confirm this result by
performing a Schr\"odinger potential analysis. Moreover we comment on the
expected form of the condensate, the explicit calculation of which we leave for the
future.

\section{Holographic Setup}
\label{sec:Holographic-Setup}
We consider $SU(2)$ Einstein-Yang-Mills theory in $(4+1)$-dimensional asymptotically AdS space. The action is
\begin{equation}
\label{eq:action}
S=\int\!\dd^5 x\sqrt{-g}\left[\frac{1}{2\kappa^2_5}\left(R-2\Lambda\right)-\frac{1}{4\hat g^2}F^a_{\mu\nu}F^{a\mu\nu}\right]+S_{\text{bdy}}~,
\end{equation}
where $\kappa_5$ is the five-dimensional gravitational constant, $\Lambda=-\frac{6}{L^2}$ is the cosmological constant, with $L$ being the AdS radius, and $\hat g$ is the Yang-Mills coupling constant. The $SU(2)$ field strength $F^a_{\mu\nu}$ is
\begin{equation}
\label{eq:fieldstrength}
F^a_{\mu\nu}=\del_\mu A^a_\nu-\del_\nu A^a_\mu+\epsilon^{abc}A^b_\mu A^c_\nu\,,
\end{equation}
where $\mu,\nu = \{t, x, y, z,u\}$, with $u$ being the AdS radial coordinate, and $\epsilon^{abc}$ is the totally antisymmetric tensor with $\epsilon^{123}=+1$. The $A^a_\mu$ are the components of the matrix-valued gauge field, $A = A^a_\mu\tau^a\dd x^\mu$, where the $\tau^a$ are the $SU(2)$ generators, which are related to the Pauli matrices by $\tau^a=\sigma^a/2\ii$.  $S_{\text{bdy}}$ includes boundary terms that do not affect the equations of motion, namely the Gibbons-Hawking boundary term as well as counterterms required for the on-shell action to be finite.

The Einstein and Yang-Mills equations derived from the above action are
\begin{align}
\label{eq:einsteinEOM}
R_{\mu \nu}+\frac{4}{L^2}g_{\mu \nu}&=\k^2\left(T_{\mu\nu}-\frac{1}{3}{T_{\rho}}^{\rho}g_{\mu\nu}\right)\,, \\
\label{eq:YangMillsEOM}
\nabla_\mu F^{a\mu\nu}&=-\epsilon^{abc}A^b_\mu F^{c\mu\nu} \,,
\end{align}
where the Yang-Mills stress-energy tensor $T_{\mu\nu}$ is
\begin{equation}
\label{eq:energymomentumtensor}
T_{\mu \nu}=\frac{1}{\hat{g}^2}{\rm tr}\left(F_{\rho\mu}{F^{\rho}}_{\nu}-\frac{1}{4}g_{\mu\nu} F_{\rho\sigma}F^{\rho\sigma}\right)\,.
\end{equation}
In order to simplify the calculation we will from now on exclusively consider the probe limit $\kappa_5/\hat g\to 0.$ Thus the metric at finite temperature is given by an AdS Schwarzschild black hole
\begin{equation}
\label{eq:metric}
\dd s^2=\frac{L^2}{u^2}\left(-f(u)\dd t^2+\dd x^2+\dd y^2+\dd z^2+\frac{\dd u^2}{f(u)}\right)\,,
\end{equation}
with blackening factor $f(u)=1-u^4/u_h^4$. The Hawking temperature of this black hole is given by $T=1/\pi u_h$. Due to the probe limit we now only have to consider the $SU(2)$ gauge fields living in this fixed background.

In this paper we introduce a constant flavour magnetic field $B=F^3_{xy}$ which solves the Yang-Mills equation of motion~\eqref{eq:YangMillsEOM}. Choosing the magnetic field in the $z$-direction breaks the rotational symmetry $SO(3)$ down to $SO(2)$ while choosing the third flavour direction breaks the $SU(2)$ symmetry down to $U(1)$ symmetry generated by $\tau^3$ rotations. We call this $U(1)$ symmetry $U(1)_3$. We may choose a gauge where only $A^3_y=xB$ is non-zero (see~\ref{sec:Constructing-gauge-covariant}).

\section{Perturbations About the Equilibrium}
\label{sec:Perturbations-about-the}
A system which is close to equilibrium can be described by linear response theory. There the effect of an external perturbation is given by the convolution of the retarded two-point function with the source of the perturbation. Due to the Cauchy integration formula, the response can be written as a sum over the poles of the two-point function (see \eg\cite{Amado:2008ji}). These poles can be identified with the different quasinormal modes  of the black hole \cite{Kovtun:2005ev}.

Quasinormal modes of a black hole are distinct perturbations of the black hole solution. They can be understood roughly as resonances of the black hole. However since the energy of the perturbation can leak into the black hole, these fluctuations are not normal modes and thus have been dubbed quasinormal. Each quasinormal mode's corresponding frequency consists of a real and an imaginary part. As for the damped oscillator, the real part of the frequency essentially determines the energy of the fluctuations, while the imaginary part is responsible for the damping. Writing the time dependence as $\ee^{-\ii\omega t}$ we note that a relaxation towards equilibrium can only happen if all the quasinormal modes lie in the lower complex half plane. In AdS spacetimes the quasinormal modes satisfy the following boundary conditions: at the horizon they are purely ingoing, whereas at the conformal AdS boundary they have an asymptotic behaviour that corresponds to a normalisable mode.

Let us now study perturbations of the gauge fields $a_\mu^a$ about the equilibrium given by the magnetic field $A_y^3$.  As usual we fix a gauge by setting $a^a_u=0$, which however leaves some residual gauge freedom. It is useful to consider combinations of the fluctuations $a_\mu^a$ which do not transform under the residual gauge transformations. These combinations are the physical modes of the system. First we define $E^\pm_\mu=a^1_\mu\pm\ii a^2_\mu$ which are in the fundamental representation of $U(1)_3$ while $a^3_\mu$ are of course in the adjoint (see \eg \cite{Erdmenger:2007ja}). Since fluctuations $a^3_\mu$ cannot interact with the magnetic field, their equations of motion are the same as for zero magnetic field. Thus we cannot observe any instability there and we will consistently set $a^3_\mu=0$. We can also consistently choose that the fluctuations have a dependence only on $t$, $x$ and $u$.

Now we combine the fields $E^{\pm}$ in such a way that they are covariant under the residual gauge symmetries, \ie they only transform as a fundamental field. Using the Fourier ansatz $\ee^{-\ii\omega t}$, this leads to the six gauge covariant fields, (see ~\ref{sec:Constructing-gauge-covariant})
\begin{equation}
\label{eq:invariantfields}
E^\pm_{L,1}=x^2B^2 E^\pm_x\pm\ii \left(xB\del_x E^\pm_y-B E^\pm_y\right)\,,\: E^\pm_{L,2}=\pm xB E^\pm_t+\omega E^\pm_y\,,\: E^\pm_T=E_z^\pm\,.
\end{equation}

The equations of motion for $E^+_{t,x,y}$, as derived from~\eqref{eq:YangMillsEOM}, are given by the coupled set of equations
\begin{align}
0=&\del_u^2 E^+_t-\frac{1}{u}\del_u E^+_t-\frac{(Bx)^2}{f}E^+_t+\frac{1}{f}\del_x^2 E^+_t-\frac{\omega Bx}{f} E^+_y+\frac{\ii\omega}{f}\del_x E^+_x\,,\label{eq:Et}\\
0=&\del_u^2 E^+_x+\left(-\frac{1}{u}+\frac{f'}{f}\right)\del_u E^+_x-\frac{(Bx)^2}{f}E^+_x+\frac{\ii B}{f}(1-x\del_x)E^+_y\nonumber\\
	&-\frac{\ii\omega}{f^2}\del_x E^+_t+\frac{\omega^2}{f^2}E^+_x\,,\label{eq:Ex}\\
0=&\del_u^2 E^+_y+\left(-\frac{1}{u}+\frac{f'}{f}\right)\del_u E^+_y-\frac{\ii B}{f}(2+x\del_x)E^+_x+\frac{1}{f}\del_x^2 E^+_y\nonumber\\
	&+\frac{Bx\omega}{f^2}E^+_t+\frac{\omega^2}{f^2}E^+_y\,,\label{eq:Ey}\\
0=&-\ii Bxf\del_u E^+_y-f\del_x\del_u E^+_x-\ii\omega\del_u E^+_t\,,\label{eq:Econst}
\end{align}
where the equations of motion for $E^-_{t,x,y}$ are obtained by complex conjugation.  Notice that it is consistent to switch off $E^\pm_t$ since \eqref{eq:Et} then coincides with the constraint \eqref{eq:Econst} and the gauge covariant fields $E^\pm_{L,2}$ disappear. The equations of motion for $E^\pm_z$ decouple from the above system since they are the only transverse modes and thus can be studied on their own.

\section{Finding the Instability}
\label{sec:Finding-the-instability}
In order to find the instabilities of the system we determine the spectrum of the quasinormal modes and determine the critical value of the magnetic field $B$ for which the imaginary part of the quasinormal modes becomes positive. Usually the crossing to the upper complex half plane occurs at the origin (see \eg \cite{Gubser:2008wv,Erdmenger:2008yj}) which we also expect here. By considering the equations of motion \eqref{eq:Et}--\eqref{eq:Econst}, we observe that $E_t^\pm$ decouple if $\omega=0$. Thus we can look for normalisable modes of these fluctuations for $\omega=0$ separately. Since we cannot find a normalisable mode for any magnetic field, we can conclude that the fluctuations $E_t^\pm$ do not trigger any instability. Therefore we consider from now on $E_t^\pm=0$ which is consistent.

Also, the equations of motion for $E^\pm_z$ decouple from all the other equations of motion since they are the transverse modes. We numerically checked that these fluctuations do not introduce an instability either. Thus an instability can only come from the $E^\pm_{x,y}$ components, the quasinormal modes of which are studied in the following.

\subsection{Analysing the Quasinormal Modes}

For $E_t^\pm\equiv 0$ it can be shown that the set of equations~\eqref{eq:Et}--\eqref{eq:Econst} reduces to one equation of motion for $E^+_x$,
\begin{equation}
\label{eq:eomEx}
0=\del_u^2E^+_x+\frac{1}{f}\del_x^2 E_x^+ +\left(\frac{f'}{f}-\frac{1}{u}\right)\del_u E_x^+ -\frac{2}{xf}\del_x E_x^+ +\left(\frac{\omega^2}{f^2}-\frac{B^2x^2}{f}\right)E_x^+\,,
\end{equation}
while $E^+_y$ is given by the constraint
\begin{equation}
\label{eq:constEy}
E_y^+=\frac{\ii}{Bx}\del_x E^+_x\,.
\end{equation}
Usually it is favourable to write down the equations of motion in terms of the gauge covariant fields~\eqref{eq:invariantfields}. That is also possible in this case, but here we obtain the solution for the gauge covariant field in terms of the solution for $E_x^+$ by using the constraint~\eqref{eq:constEy},
\begin{equation}
\label{eq:EL1Ex}
E^\pm_{L,1}=-\left(\del_x^2-\frac{2}{x}\del_x-x^2B^2\right) E^\pm_x\,.
\end{equation}
From this it is clear that the dependence on the radial coordinate $u$ is the same for both fields, and only the dependence on $x$ may change.

The solution for $E_x^+$ can be found by the separation of variables $E_x^+=X(x)U(u)$ and its equation of motion~\eqref{eq:eomEx} becomes
\begin{equation}
\label{eq:eqseparation}
\begin{split}
0&=X''-\frac{2}{x}X'-\left(C+B^2x^2\right)X\,,\\
0&=U''+\left(\frac{f'}{f}-\frac{1}{u}\right)U'+\left(\frac{\omega^2}{f^2}+\frac{C}{f}\right)U\,,
\end{split}
\end{equation}
where $C$ is the separation constant. Using~\eqref{eq:EL1Ex}, the equation for $X$ and the fact that $E^\pm_x$ is separable, we find that
\begin{equation}
 E^\pm_{L,1}= -C E^\pm_x~.
\end{equation}
This shows that the physical fields $E^\pm_{L,1}$ and $E^{\pm}_x$ have the same $x$-dependence and thus are equivalent up to a constant. In solving for $X$, we will have to impose boundary conditions that ensure that the $E^\pm_{L,1}$ vanish asymptotically and are regular. As we will find out, they have an overall factor of $\exp(-B x^2/2)$.

The equation for $X$ can be solved analytically. After the definition $X=\exp(-B x^2/2)Y$ and $y=Bx^2$ we obtain Kummer's equation
\begin{equation}
\label{eq:}
y Y''(y)+(b-y)Y'(y)-aY(y)=0\,,
\end{equation}
with $b=-1/2$ and $a=(C-B)/4B$. The solution is given by the linear combination of the confluent hypergeometric function of first kind $\M(a,b,y)$ and of the second kind $\U(a,b,y)$. The function $\U$ can also be written in terms of $\M$ such that the complete solution is given by the linear combination
\begin{equation}
\label{eq:solY}
Y=C_1 \M(a,b,y)+C_2 y^{1-b}\M(1+a-b,2-b,y)\,.
\end{equation}
In order to have a vanishing solution for $E^\pm_{L,1}$ (and thus for $X$) at $x\to\infty$, $Y$ is only allowed to grow slower than the factor of $\exp(y/2)$ since $X=\exp(-y/2)Y$. Looking at the asymptotic expansion for large $y$,
\begin{equation}
 \M(a;b;y)
 \sim
 e^{y} y^{a - b} \left(\frac{\Gamma(b)}{\Gamma(a)}
 +
 \mathcal{O}\left(\frac{1}{y}\right)
 \right)
 + \frac{(-y)^{-a} \Gamma(b)}{\Gamma(b-a)} \left(
 1
 +
 \mathcal{O}\left(\frac{1}{y}\right)
 \right)~,
\end{equation}
we see that $Y$ does grow too fast unless $\Gamma(a)$ diverges, which is to say that $a=-m$ for a non-negative integer $m$. Indeed, when $a=-m$ then $\M(a,b,y)$ reduces to a Laguerre polynomial $L^{(\alpha)}_m(y)$, which has $m$ positive roots. We expect that the free energy of the system increases with the number of the roots. Since we are looking for the lowest energy solution and the second solution in~\eqref{eq:solY} always has a root at $y=0$, we obtain $C_2=0$. The first solution has no zero if $a=0$. This determines
\begin{equation}
\label{eq:solX}
C=B\text{ and }X=-\ee^{-Bx^2/2}\,,
\end{equation}
where the normalisation is set by $X(0)=-1$. We thus find that
\begin{equation}
 E^\pm_{L,1}= B\ee^{-Bx^2/2} U(u) ~.
\end{equation}

\begin{figure}
\psfrag{imw}{$\mathrm{Im}~\mathfrak{w}$}
\psfrag{rew}{$\mathrm{Re}~\mathfrak{w}$}
\centering
\includegraphics[width=0.8\textwidth]{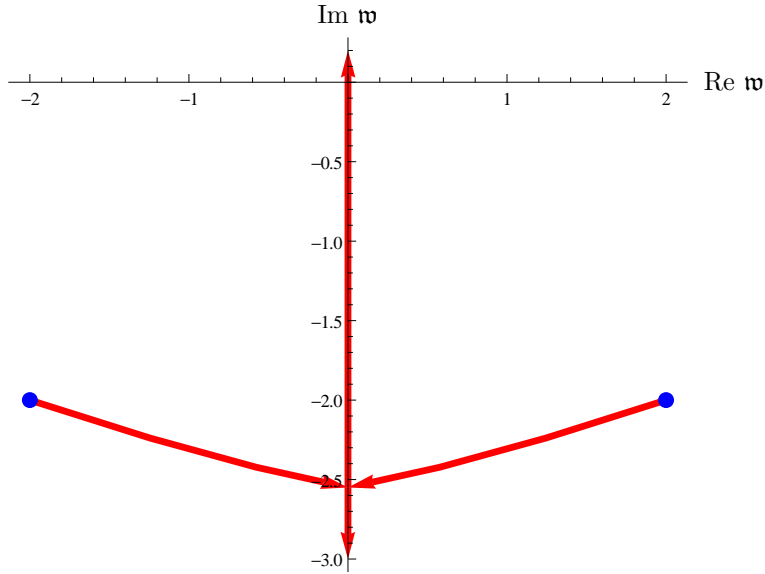}
\caption{The movement of the quasinormal frequencies at increasing
$\mathfrak{B}$. The two lowest order frequencies 
start at $\mathfrak{w}=\pm 2 - 2 i$. They move towards the imaginary axis,
merge, then split again. The merging happens when $\mathfrak{B}\approx 1.92$.
One pole then moves downward, away from 0 and outside the range of the figure.
The other, which is drawn, moves towards 0 and eventually crosses the real axis
when $\mathfrak{B}=\mathfrak{B}_c \approx 5.15$.}
\label{fig:polemovement}
\end{figure}

Unfortunately the equation for $U$ can only be solved numerically. Using a shooting method we may find its quasinormal modes and determine the dependence of their frequencies on the magnetic field. In order to have dimensionless quantities we define $\mathfrak{B}=Bu_h^2=B/(\pi T)^2$ and $\mathfrak{w}=\omega u_h=\omega/\pi T$. For $\mathfrak{B}=0$ the quasinormal frequencies are given by $\mathfrak{w}_n=n(\pm 2-2i)$ \cite{Nunez:2003eq}. The dependence of the quasinormal frequencies on the magnetic field is given in figure~\ref{fig:polemovement}. We observe that for the magnetic field $\mathfrak{B}\approx 5.15$, the quasinormal frequency moves into the upper half plane, which signals an instability.

\subsection{Tachyonic Mode}
There are two further ways of qualitatively understanding the formation of an instability that leads to a superconducting condensate. The first is by looking at the form that the on-shell action takes in our setup. To see an instability we actually only need to switch on the fields $A^3_y=x B$, $a^1_x$ and $a^2_y$. To second order, the relevant part of the action becomes
\begin{equation}
\begin{split}  S_{(2)}=
 -\frac{1}{4\hat g^2} \int\!\dd^5 x\sqrt{-g} \bigg\{&
   2 g^{tt} g^{xx} \left(\partial_t a^1_x\right)^2
   + 2 g^{tt} g^{yy} \left(\partial_t a^2_y\right)^2+ 2 g^{uu} g^{xx} \left(\partial_u a^1_x\right)^2\\
   &+ 2 g^{uu} g^{yy} \left(\partial_u a^2_y\right)^2+ 2 g^{xx} g^{yy} \Big[
     B^2 + \left(\partial_x a^2_y \right)^2\\
     & + 2B a^1_x a^2_y
     - 2 B x a^1_x \partial_x a^2_y + B^2 x^2 \left(a^1_x\right)^2
   \Big]
 \bigg\}\,.
\end{split}
\end{equation}
Solving the $x$-dependence, we obtain $E^+_x = \ee^{-\frac{1}{2} B x^2} V(t,u)$ and equation~\eqref{eq:constEy} gives $E^+_y=-i E^+_x$ which translates to $a^1_x = -a^2_y = \ee^{-\frac{1}{2} B x^2} \re V$. This means that, to second order, the on-shell action is
\begin{equation}
\begin{split}  S_{(2)}=
 -\frac{1}{4\hat g^2} \int\! \dd t \dd y \dd z \dd u \bigg[&
   2 g^{tt} \left( g^{xx} + g^{yy} \right) \left(\partial_t \re V\right)^2\\
 &+ 2 g^{uu} \left( g^{xx} + g^{yy} \right) \left(\partial_u \re V \right)^2
   - 4 B g^{xx} g^{yy} \left(\re V\right)^2
 \bigg]\,,\end{split}
\end{equation}
where we have integrated out the $x$ component, absorbed a factor of
$\left(\pi/B\right)^\frac{1}{4}$ into $V$ for the canonical
normalisation and dropped a constant term. The important thing to
notice is that the gauge field has picked up an effective mass term,
$m^2_\mathrm{eff} = - 4 B g^{xx} g^{yy}$, and it becomes more and more
negative as $B$ increases. This suggests a tachyonic mode leading to
the spontaneous formation of a condensate. In contrast to the
instability at finite chemical potential \cite{Gubser:2008wv}, the
instability here does not arise due to the presence of a horizon, \ie $g^{tt}$ is not involved in the negative effective mass term.

\subsection{Schr\"odinger Potential Analysis}
\begin{figure}
\centering
\includegraphics[width=0.8\textwidth]{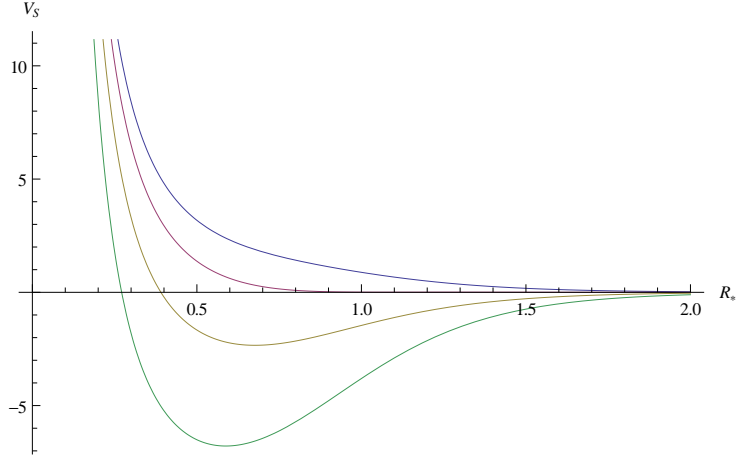}
\caption{The Schr\"odinger potential at different values of $\mathfrak{B}$.
It is helpful to compare this with figure~\ref{fig:polemovement}.
From top to bottom, the curves show the potential at the values
$\mathfrak{B}=0,~1.92,~5.15$ and $10$. The first curve shows the potential when
there is no magnetic field, the second
one is for the value of $\mathfrak{B}$ at which the poles merge on the imaginary
axis, the third
is for the value at which the first pole crosses the real axis, and the final
value
is to see what the potential does at higher $\mathfrak{B}$. We see that a
potential well forms with negative potential energy. The value
$\mathfrak{B}=1.92$ is roughly where the curve first dips below zero, which is
where the poles in figure~\ref{fig:polemovement} merge.}
\label{fig:schroedpotential}
\end{figure}

To understand the instability better, we can rewrite the equation of motion \eqref{eq:eomEx} in the form of a Schr\"odinger equation (see \eg \cite{Hoyos:2006gb,Myers:2007we}),
\begin{equation}
\label{eq:Schroedingereq}
-\del_{R}^2\psi+V_s\psi=E\psi,
\end{equation}
where $E=\mathfrak{w}^2$ and $\psi\propto E_x^+$. The tortoise coordinate $R$ and the Schr\"odinger potential are given by
\begin{equation}
\label{eq:Randpotential}
R=\frac{1}{2}(\operatorname{arctanh}u+\arctan u)\quad\text{and}\quad V_s=\frac{(1-u^4)(3-4Bu^2+5u^4)}{4u^2}\,.
\end{equation}
$R(u)$ cannot be inverted analytically and so we have to study $V_s(R)$ numerically. The profile of the Schr\"odinger potential for different values of the magnetic field $\mathfrak{B}$ is given in figure~\ref{fig:schroedpotential}. For all values of the magnetic field we observe an infinite wall at $R=0$ due to the boundary of the AdS space and a vanishing potential at $R\to\infty$ due to the black hole horizon. In between a potential well opens up as the magnetic field is increased. For a large enough magnetic field, this potential allows for bound states with negative energy. The first such bound state implies that a quasinormal frequency has crossed the real line and developed a positive imaginary part. This bound state therefore first forms at the critical magnetic field that produces an instability.

The reason why a bound state with negative energy implies an instability is due to the ingoing boundary condition at the horizon. This boundary condition implies that the time-independent part of the mode goes like $\ee^{\ii\omega R}$. Outside the potential well, the mode has to decay exponentially towards the horizon. Combining these two constraints means that $\omega$ must have a positive imaginary part. Thus the appearance of the bound state in the Schr\"odinger picture agrees with the quasinormal frequency crossing the real axis. 

Using the Schr\"odinger potential we can also estimate the critical B-field $\mathfrak{B}_c,$ above which the system gets unstable. Using a WKB approximation, we can estimate that a (zero-energy) bound state will appear for
\begin{equation}
\label{eq:WKB}
\left( n-\frac{1}{2} \right) \pi =\int_{R_0}^\infty\! \dd R\, \sqrt{-V_{s}(R)}=\int_{u_0}^1\! \frac{\dd u}{1-u^4} \sqrt{-V_{s}(u)} \, , 
\end{equation}
where $n \in \mathbbm{N}.$ $R_0$ is the zero of $V_s(R)$ and thus $V_s(R)$ is
negative for $R>R_0.$ The corresponding zero of $V_s$ as a function of $u$ is
called $u_0,$
\begin{equation}
u_0 = \sqrt{\frac{2 \mathfrak{B}}{5}-\frac{1}{5} \sqrt{4 \mathfrak{B}^2-15}}\,.
\end{equation}
Note that $V_s(u)$ and therefore also $u_0$ depend on $\mathfrak{B}.$ The minimal value for $\mathfrak{B}$ where the potential develops a zero is at $\mathfrak{B}=\frac{\sqrt{15}}{2}\approx 1.94$. This coincides with the value at which the quasinormal modes merge at the negative imaginary axis up to an error of 2\%. 

In order to obtain the critical magnetic field we need to know when a bound state can be formed. We are only interested in the first bound state and therefor set $n=1.$ Solving the integral equation \eqref{eq:WKB} numerically for $\mathfrak{B}$ we obtain $\mathfrak{B}_c \approx 5.3514.$  We see that the analysis of the Schr\"odinger potential qualitatively agrees with the precise study of the quasinormal modes above up to an error of 4\%.

\section{Conclusion}
\label{sec:Conclusion}
In this letter we have found an interesting instability of black holes in asymptotically AdS spacetimes. In particular we consider an Einstein-Yang-Mills system with gauge group $SU(2)$ and with a negative cosmological constant such that AdS is a solution. If we neglect the back-reaction of the gauge fields on the background metric, a solution of the equations of motion is given by an asymptotically Schwarzschild black hole.

In addition we turn on a magnetic field $F_{xy} = B \tau^3,$ where $\tau^3$ is one of the three generators of $SU(2)$. By investigating the fluctuations of the gauge field around the Schwarzschild black hole we find that the system gets unstable above a critical value $\mathfrak{B}_c \approx 5.15$ of the magnetic field. In particular we identify the quasinormal mode which crosses the real axis into the upper half plane. This result is confirmed in two different ways. First, we calculate an effective action for the fluctuations and show that these fluctuations get a negative mass squared, which drops below the Breitenlohner-Freedman bound for large enough magnetic fields. Second, we rewrite the equations of motion into a Schr\"odinger-like wave equation with an effective potential $V_s$ and show that $V_s$ has a bound state for large enough magnetic fields. The bound state signals the instability of the system.

Note that here we have turned on a magnetic field for one of the three
generators of $SU(2)$ and not for a $U(1)$ gauge group. One might think
that this is not so common in gauge/gravity duality.
However, as we now discuss, a similar structure \cite{Callebaut:2011uc,Callebaut:2011ab} also arises in the
Sakai-Sugimoto model \cite{Sakai:2004cn}. 
To see this, 
let us consider the Sakai-Sugimoto setup with two flavours (with
electric charges $q_1$ and $q_2$). The flavour degrees of freedom are
described by two pairs of $D8/\overline{D8}$-branes. Part of the
dynamics of the flavour degrees of freedom is therefore given by   a
non-Abelian gauge theory with gauge group $U(2)_L \times U(2)_R.$ Note
that the gauge group on the gravity side 
corresponds to the dual flavour symmetry on the field theory side. In the Sakai-Sugimoto model the electromagnetic gauge group with generator $Q=\diag(q_1,q_2)$ is realised as a subgroup of $U(2)_L \times U(2)_R.$ Thus in the case $q_1 \neq q_2,$ the magnetic field $\mathcal{B}_{em}$ has a baryon and isospin component, i.e.
\begin{equation}
Q \mathcal{B}_{em} = \frac{q_1 +q_2}{2}  \mathcal{B}_{em} \mathbbm{1} +
\frac{q_1-q_2}{2}  \mathcal{B}_{em} \tau_3.
\end{equation}
Although we aim to study the effect of the $U(1)$ magnetic field
$\mathcal{B}_{em},$ we have to consider a non-vanishing magnetic field
for the isospin component. Note that already in
\cite{Callebaut:2011uc,Callebaut:2011ab}, evidence was given that the
vacuum at zero temperature is unstable for large enough magnetic
fields. The results presented here indicate that such an instability
arises not only in the Sakai-Sugimoto model but also in the simplest
possible toy model at finite temperature for which we can explicitly study the quasinormal modes. 

The instability discussed in this paper gives rise to interesting
questions on which we plan to work in the future. What is the true
ground state of the system and what are its properties? Is it a
superfluid state? Can we realise this ad hoc toy model in string
theory? By addressing these questions, we expect in particular to
obtain further information about the QCD vacuum instability at finite
$B$ field discussed in \cite{Chernodub:2010qx,Chernodub:2010zw,Chernodub:2011mc,Chernodub:2011tv}. As our results suggest, a more
general form of this instability may occur also at finite temperature
in non-confining field theories. On the gravity side, we expect the
condensate to be a suitable combination of the gauge field components 
$A_x^1$ and $A_y^2$. 

It will also be interesting to repeat the analysis of this paper in
four dimensions on the gravity side. This will allow for a comparison
with  
\cite{Mosaffa:2006gp},  where instabilities of general spherically symmetric
solutions of four-dimensional Einstein-Yang-Mills-Dilaton theories
with a constant magnetic field strength on the sphere have been
found. Moreover, this will also allow for a comparison with 
\cite{Sutcliffe:2011sr}, which
raises the interesting question whether the instability found here is
related to the existence of magnetic monopoles. 

\section*{Acknowledgments}\label{sec:Acknowledgments}
We would like to thank Nele Callebaut, 
Maxim Chernodub, David Dudal and Henri Verschelde
 for useful conversations. The work of M.A. is supported by NSF grant PHY-07-57702. The work of J.E., P.K. and M.S. is supported in part by \textit{The Cluster of Excellence for Fundamental Physics - Origin and
Structure of the Universe}. 

\appendix
\section{Constructing the Gauge Covariant Fields}\label{sec:Constructing-gauge-covariant}
In this appendix we show how to fix the gauge of the $SU(2)$ field in such a way that only a $U(1)_3$ gauge freedom remains. We then find the fluctuations that transform covariantly under the remaining $U(1)_3$.

For the background field $A^a_\mu$ without fluctuations, the gauge transformation $e^{i \Lambda(x^\mu)}$ acts infinitesimally on $A^a_\mu$ as
\begin{equation}
 \delta A^a_\mu = \partial_\mu \Lambda^a - \epsilon^{abc} \Lambda^b A^c_\mu~.
\end{equation}
As usual we fix the gauge such that $A_u=0$. This leaves $\Lambda^a(t,x,y,z)$ as the residual gauge symmetry. 
We choose to have the magnetic field $B=F^3_{xy}$. From~\eqref{eq:fieldstrength} we see that this means we can fix
$A_x^a=0$, $A^{1,2}_y=0$ and $A^3_y=xB$. These choices give that $F^a_{xy}=\del_x A^a_y=B\delta^{a3}$. The choice that $A_x^a=0$ reduces the remaining gauge symmetry to $\Lambda^a(t,y,z)$, and the conditions on $A^a_y$ fix $\Lambda^{1,2}=0$ such that only $\Lambda^3(t,z)$ is left. The $SU(2)$ has been broken to $U(1)_3$.

Now we consider the fluctuations $a_\mu^a(t,x,y,z,u)$ about the background field $A^a_\mu$. The full field $A^a_\mu+a^a_\mu$ transforms under the infinitesimal gauge transformation $\Lambda^a+\lambda^a$. We treat $\lambda^a$ as an infinitesimal perturbation of $\Lambda^a$. Then the fluctuations $a^a_\mu$ transform as
\begin{equation}
 \delta a^a_\mu=\del_\mu \lambda^a-\epsilon^{abc}(\Lambda^b a^c_\mu+\lambda^b A^c_\mu)\,,
\end{equation}
while the transformation for $A^a_\mu$ remains the same as before.
Again we fix $a_u=0$ which leaves $\lambda(t,x,y,z)$ as the residual gauge symmetry for the fluctuations. We define the fields $E^\pm=a^1\pm\ii a^2$. The gauge transformations of the fields $E^{\pm}$ and $a^3$ to first order are
\begin{equation}
\label{eq:gaugetrafofluc}
\begin{split}
\delta E^\pm_{t,x,z}&=\del_{t,x,z}\lambda^\pm\mp\ii\Lambda^3 E^\pm_{t,x,z}\,,\\
\delta E^\pm_y&=\del_y\lambda^\pm\mp\ii\Lambda^3 E^\pm_y\pm\ii A_y^3\lambda^\pm\,,\\
\delta a^3_\mu&=\del_\mu \lambda^3\,,
\end{split}
\end{equation}
where $\lambda^\pm=\lambda^1\pm\ii\lambda^2$. Notice that the fields $E^\pm_\mu$ transform in the fundamental of the $U(1)_3$. We now consider only a dependence of $t$ and $x$ and use the Fourier ansatz $\ee^{-\ii\omega t}$. The fields transforming covariantly under the $U(1)_3$ gauge symmetry are then
\begin{equation}
\label{eq:invariantfields2}
\begin{aligned}
E^\pm_{L,1}&=x^2B^2 E^\pm_x\pm\ii \left(xB\del_x E^\pm_y-B E^\pm_y\right)\,,&E^3_L&=\del_x a^3_t+\ii\omega a^3_x\,,\\
E^\pm_{L,2}&=\pm xB E^\pm_t+\omega E^\pm_y\,,&E^3_{T,1}&=a^3_y\,,\\
E^\pm_T&=E_z^\pm\,,&E^3_{T,2}&=a^3_z\,.
\end{aligned}
\end{equation}
Since the $B$ field does not interact with fields in flavour direction $3$, these $U(1)_3$ covariant fields are trivial. However the fields in direction $1$ and $2$ couple and take a non-trivial form.


\begin{thebibliography}{10}


\bibitem{Gubser:2008zu}
S.~S. Gubser, \emph{ {Colorful horizons with charge in anti-de Sitter space}},
  Phys. Rev. Lett. {\bf 101} (2008) 191601,
\href{http://www.slac.stanford.edu/spires/find/hep/www?texkey=Gubser:2008zu}{a%
rXiv:0803.3483}.

\bibitem{Gubser:2008px}
S.~S. Gubser, \emph{ {Breaking an Abelian gauge symmetry near a black hole
  horizon}}, Phys. Rev. {\bf D78} (2008) 065034,
\href{http://www.slac.stanford.edu/spires/find/hep/www?texkey=Gubser:2008px}{a%
rXiv:0801.2977}.

\bibitem{Hartnoll:2008vx}
S.~A. Hartnoll, C.~P. Herzog, and G.~T. Horowitz, \emph{ {Building a
  Holographic Superconductor}}, Phys. Rev. Lett. {\bf 101} (2008) 031601,
\href{http://www.slac.stanford.edu/spires/find/hep/www?texkey=Hartnoll:2008vx}%
{arXiv:0803.3295}.

\bibitem{Hartnoll:2008kx}
S.~A. Hartnoll, C.~P. Herzog, and G.~T. Horowitz, \emph{ {Holographic
  Superconductors}}, JHEP {\bf 12} (2008) 015,
\href{http://www.slac.stanford.edu/spires/find/hep/www?texkey=Hartnoll:2008kx}%
{arXiv:0810.1563}.

\bibitem{Gubser:2008wv}
S.~S. Gubser and S.~S. Pufu, \emph{ {The gravity dual of a p-wave
  superconductor}}, JHEP {\bf 11} (2008) 033,
\href{http://www.slac.stanford.edu/spires/find/hep/www?texkey=Gubser:2008wv}{a%
rXiv:0805.2960}.

\bibitem{Chernodub:2010qx}
M.~Chernodub, \emph{ {Superconductivity of QCD vacuum in strong magnetic
  field}}, Phys.Rev. {\bf D82} (2010) 085011,
  \href{http://www.slac.stanford.edu/spires/find/hep/www?texkey=Chernodub:2010%
qx}{arXiv:1008.1055}.

\bibitem{Chernodub:2010zw}
M.~Chernodub, \emph{ {Electromagnetically superconducting phase of QCD vacuum
  induced by strong magnetic field}}, AIP Conf.Proc. {\bf 1343} (2011)
  149--151,
  \href{http://www.slac.stanford.edu/spires/find/hep/www?texkey=Chernodub:2010%
zw}{arXiv:1011.2658}.

\bibitem{Chernodub:2011mc}
M.~Chernodub, \emph{ {Spontaneous electromagnetic superconductivity of vacuum
  in strong magnetic field: evidence from the Nambu--Jona-Lasinio model}},
  Phys.Rev.Lett. {\bf 106} (2011) 142003,
  \href{http://www.slac.stanford.edu/spires/find/hep/www?texkey=Chernodub:2011%
mc}{arXiv:1101.0117}.

\bibitem{Chernodub:2011tv}
M.~Chernodub, \emph{ {Can nothing be a superconductor and a superfluid?}},
  \href{http://www.slac.stanford.edu/spires/find/hep/www?texkey=Chernodub:2011%
tv}{arXiv:1104.4404}.

\bibitem{Nielsen:1978rm}
  N.~K.~Nielsen, P.~Olesen,
  \emph{ {An Unstable Yang-Mills Field Mode}},
  Nucl.\ Phys.\  {\bf B144 } (1978)  376.

\bibitem{Ambjorn:1988tm}
  J.~Ambjorn, P.~Olesen,
  \emph{ {On Electroweak Magnetism}},
  Nucl.\ Phys.\  {\bf B315 } (1989)  606.

\bibitem{Ambjorn:1989bd}
  J.~Ambjorn, P.~Olesen,
  \emph{ {A Condensate Solution Of The Electroweak Theory Which Interpolates
Between The Broken And The Symmetric Phase}},
  Nucl.\ Phys.\  {\bf B330 } (1990)  193.

\bibitem{Ambjorn:1989sz}
  J.~Ambjorn, P.~Olesen,
  \emph{ {Electroweak Magnetism: Theory And Application}},
  Int.\ J.\ Mod.\ Phys.\  {\bf A5 } (1990)  4525-4558.

\bibitem{Callebaut:2011uc}
N.~Callebaut, D.~Dudal, and H.~Verschelde, \emph{ {Holographic study of rho
  meson mass in an external magnetic field: Paving the road towards a
  magnetically induced superconducting QCD vacuum?}},
  \href{http://www.slac.stanford.edu/spires/find/hep/www?texkey=Callebaut:2011%
uc}{arXiv:1102.3103}.

\bibitem{Callebaut:2011ab}
N.~Callebaut, D.~Dudal, and H.~Verschelde, \emph{ {Holographic rho mesons in an
  external magnetic field}},
  \href{http://www.slac.stanford.edu/spires/find/hep/www?texkey=Callebaut:2011%
ab}{arXiv:1105.2217}.

\bibitem{Ammon:2008fc}
M.~Ammon, J.~Erdmenger, M.~Kaminski, and P.~Kerner, \emph{ {Superconductivity
  from gauge/gravity duality with flavor}}, Phys. Lett. {\bf B680} (2009)
  516--520,
\href{http://www.slac.stanford.edu/spires/find/hep/www?texkey=Ammon:2008fc}{ar%
Xiv:0810.2316}.

\bibitem{Basu:2008bh}
P.~Basu, J.~He, A.~Mukherjee, and H.-H. Shieh, \emph{ {Superconductivity from
  D3/D7: Holographic Pion Superfluid}}, JHEP {\bf 11} (2009) 070,
\href{http://www.slac.stanford.edu/spires/find/hep/www?texkey=Basu:2008bh}{arX%
iv:0810.3970}.

\bibitem{Ammon:2009fe}
M.~Ammon, J.~Erdmenger, M.~Kaminski, and P.~Kerner, \emph{ {Flavor
  Superconductivity from Gauge/Gravity Duality}}, JHEP {\bf 10} (2009) 067,
\href{http://www.slac.stanford.edu/spires/find/hep/www?texkey=Ammon:2009fe}{ar%
Xiv:0903.1864}.

\bibitem{Basu:2009vv}
P.~Basu, J.~He, A.~Mukherjee, and H.-H. Shieh, \emph{ {Hard-gapped Holographic
  Superconductors}}, Phys. Lett. {\bf B689} (2010) 45--50,
\href{http://www.slac.stanford.edu/spires/find/hep/www?texkey=Basu:2009vv}{arX%
iv:0911.4999}.

\bibitem{Ammon:2009xh}
M.~Ammon, J.~Erdmenger, V.~Grass, P.~Kerner, and A.~O'Bannon, \emph{ {On
  Holographic p-wave Superfluids with Back-reaction}}, Phys. Lett. {\bf B686}
  (2010) 192--198,
\href{http://www.slac.stanford.edu/spires/find/hep/www?texkey=Ammon:2009xh}{ar%
Xiv:0912.3515}.

\bibitem{Amado:2008ji}
I.~Amado, C.~Hoyos-Badajoz, K.~Landsteiner, and S.~Montero, \emph{
  {Hydrodynamics and beyond in the strongly coupled N=4 plasma}}, JHEP (2008)
  133,
\href{http://www.slac.stanford.edu/spires/find/hep/www?texkey=Amado:2008ji}{ar%
Xiv:0805.2570}.

\bibitem{Kovtun:2005ev}
P.~K. Kovtun and A.~O. Starinets, \emph{ Quasinormal modes and holography},
  Phys. Rev. {\bf D72} (2005) 086009,
\href{http://www.slac.stanford.edu/spires/find/hep/www?texkey=Kovtun:2005ev}{a%
rXiv:0506184}.

\bibitem{Erdmenger:2007ja}
J.~Erdmenger, M.~Kaminski, and F.~Rust, \emph{ {Holographic vector mesons from
  spectral functions at finite baryon or isospin density}}, Phys. Rev. {\bf
  D77} (2008) 046005,
\href{http://www.slac.stanford.edu/spires/find/hep/www?texkey=Erdmenger:2007ja%
}{arXiv:0710.0334}.

\bibitem{Erdmenger:2008yj}
J.~Erdmenger, M.~Kaminski, P.~Kerner, and F.~Rust, \emph{ {Finite baryon and
  isospin chemical potential in AdS/CFT with flavor}}, JHEP {\bf 11} (2008)
  031,
\href{http://www.slac.stanford.edu/spires/find/hep/www?texkey=Erdmenger:2008yj%
}{arXiv:0807.2663}.

\bibitem{Nunez:2003eq}
A.~Nunez and A.~O. Starinets, \emph{ {AdS/CFT correspondence, quasinormal
  modes, and thermal correlators in N = 4 SYM}}, Phys. Rev. {\bf D67} (2003)
  124013,
\href{http://www.slac.stanford.edu/spires/find/hep/www?texkey=Nunez:2003eq}{ar%
Xiv:0302026}.

\bibitem{Hoyos:2006gb}
C.~Hoyos-Badajoz, K.~Landsteiner, and S.~Montero, \emph{ {Holographic Meson
  Melting}}, JHEP {\bf 04} (2007) 031,
\href{http://www.slac.stanford.edu/spires/find/hep/www?texkey=Hoyos:2006gb}{ar%
Xiv:0612169}.

\bibitem{Myers:2007we}
R.~C. Myers, A.~O. Starinets, and R.~M. Thomson, \emph{ {Holographic spectral
  functions and diffusion constants for fundamental matter}}, JHEP {\bf 11}
  (2007) 091,
\href{http://www.slac.stanford.edu/spires/find/hep/www?texkey=Myers:2007we}{ar%
Xiv:0706.0162}.

\bibitem{Sakai:2004cn}
T.~Sakai and S.~Sugimoto, \emph{ {Low energy hadron physics in holographic
  QCD}}, Prog.Theor.Phys. {\bf 113} (2005) 843--882,
  \href{http://www.slac.stanford.edu/spires/find/hep/www?texkey=Sakai:2004cn}{%
arXiv:0412141}.

\bibitem{Mosaffa:2006gp}
  A.~E.~Mosaffa, S.~Randjbar-Daemi, M.~M.~Sheikh-Jabbari,
  \emph{ {Non-Abelian magnetized blackholes and unstable attractors}},
  Nucl.\ Phys.\  {\bf B789 } (2008)  225-244,
\href{http://www.slac.stanford.edu/spires/find/hep/www?texkey=Mosaffa:2006gp}{
arXiv:0612181}.

\bibitem{Sutcliffe:2011sr}
  P.~Sutcliffe,
  \emph{ {Monopoles in AdS}},
\href{http://www.slac.stanford.edu/spires/find/hep/www?texkey=Sutcliffe:2011sr}{
arXiv:1104.1888}.

\end{thebibliography}
\providecommand{\href}[2]{#2}\begingroup\raggedright\endgroup

\end{document}